\documentclass[aip,preprint]{revtex4-1}

\usepackage{graphicx}
\usepackage{subcaption}
\usepackage{dcolumn}
\usepackage{bm}
\usepackage{hyperref}
\usepackage{url}
\usepackage[usenames, dvipsnames]{color}
\usepackage{color}

\draft 

\begin{document}

\title{Coincidence velocity map imaging using Tpx3Cam, a time stamping optical camera with 1.5 ns timing resolution} 



\author{Arthur Zhao}
\affiliation{Department of Physics and Astronomy, Stony Brook University, Stony Brook NY 11794-3800}
\author{Martin van Beuzekom}
\affiliation{Nikhef, Science Park 105, 1098 XG Amsterdam, Netherlands}
\author{Bram Bouwens}
\affiliation{Amsterdam Scientific Instruments, Science Park 105, 1098 XG Amsterdam, Netherlands}
\author{Dmitry Byelov}
\affiliation{Amsterdam Scientific Instruments, Science Park 105, 1098 XG Amsterdam, Netherlands}
\author{Irakli Chakaberia}
\affiliation{Brookhaven National Laboratory, Upton NY 11973}
\affiliation{Kent State University, Kent OH, 44242}
\affiliation{Shandong University, Jinan, Shandong 250100, China}
\author{Chuan Cheng}
\affiliation{Department of Physics and Astronomy, Stony Brook University, Stony Brook NY 11794-3800}
\author{Erik Maddox}
\affiliation{Amsterdam Scientific Instruments, Science Park 105, 1098 XG Amsterdam, Netherlands}
\author{Andrei Nomerotski}
\affiliation{Brookhaven National Laboratory, Upton NY 11973}
\author{Peter Svihra}
\affiliation{Faculty of Nuclear Sciences and Physical Engineering, Czech Technical University, Prague 115 19, Czech Republic}
\author{Jan Visser}
\affiliation{NIKHEF, Science Park 105, 1098 XG Amsterdam, Netherlands}
\author{Vaclav Vrba}
\affiliation{Faculty of Nuclear Sciences and Physical Engineering, Czech Technical University, Prague 115 19, Czech Republic}
\author{Thomas Weinacht}
\email[Corresponding author, ]{thomas.weinacht@stonybrook.edu}
\homepage[]{http://ultrafast.physics.sunysb.edu/}
\affiliation{Department of Physics and Astronomy, Stony Brook University, Stony Brook NY 11794-3800}



\date{\today}

\begin{abstract}
	We demonstrate a coincidence velocity map imaging apparatus equipped with a novel time stamping fast optical camera, Tpx3Cam, whose high sensitivity and ns timing resolution allow for simultaneous position and time-of-flight detection. This single detector design is simple, flexible and capable of highly differential measurements. We show detailed characterization of the camera and its application in strong field ionization experiments.
\end{abstract}


\maketitle 

\section{Introduction}

Coincidence velocity map imaging (VMI)\cite{chandler1987two,eppink1997velocity,ullrich2003recoil,vredenborg2008photoelectron} has become an essential tool in the study of reaction dynamics and strong field laser-matter interactions  \cite{krausz2009attosecond,davies1999femtosecond,downie2000angle,lebech2002ion,doerner1998photo,rijs2004femtosecond,tang2009threshold,kling2014thick}. VMI maps the transverse momenta of charged particles to positions on a 2D detector such that for a given particle species its distance to the center of the detector is proportional to its initial transverse velocity. One can reconstruct the 3D momentum distribution from the 2D projection via Abel inversion, given that the initial distribution possesses a cylindrical symmetry, with the axis of symmetry parallel to the detector plane. Coincidence VMI detects both electrons and cations emitted from the same atom/molecule, allowing for access to the full kinematics of a single reaction, and for systematic studies of its underlying physical mechanisms. The most natural design for a coincidence VMI apparatus includes two sets of electrostatic lenses for VMI and two sets of time and position sensitive 2D detectors, for electron and cation detection. The detector usually consists of a set of microchannel plates (MCP), acting as an amplifier, and an anode detector, such as a delay line detector, which provides excellent time and spatial resolution. Simpler and more flexible implementations have also been introduced, where a single set of electrostatic lenses with time-dependent voltages projects both electrons and cations onto a single detector  \cite{lehmann2012velocity}. The delay line detector can be replaced by a fast camera for VMI, while the time-of-flight (ToF) information is collected via a fast digitizer coupled to the MCPs. This configuration allows for easy switching between coincidence and non-coincidence acquisition modes \cite{zhao2017coincidence}. Non-coincidence acquisition mode is particularly useful in exploring larger parameter spaces (intensity, time duration, pulse shape, pump-probe delay, etc.) and for calibration purposes.

Recent experiments have started to take advantage of new imaging technologies where each individual pixel in a sensor functions independently and is able to time stamp an incident ``event". This transforms the imaging sensor into an array of fast digitizers with both spatial and temporal resolutions. Examples of such approach include PImMS cameras, which use monolithic CMOS (complimentary metal-oxide conductor) technology\cite{nomerotski2010pimms,clark2012multimass} and direct detection of electrons and ions after an MCP employing Timepix ASIC (Application Specific Integrated Circuit)\cite{jungmann2010,heeren2012,llopart2007timepix}.

In this manuscript, we describe a coincidence VMI detector utilizing a novel time-stamping fast optical camera, Tpx3Cam. The camera has excellent spatial and temporal resolution and can simultaneously measure the mass and 3D momentum of all charged particles in each ionization event. The camera's large throughput enables a straightforward switch between low-rate coincidence and high-rate non-coincidence detection modes. In addition, as a standalone piece of equipment outside the vacuum chamber, it allows for easy upgrades and maintenance.

Below, we will first describe our experimental setup, in particular, Tpx3Cam. Then we will present new results from two coincidence experiments. These are proof-of-principle experiments, using a molecule (bromoiodomethane, $\mathrm{CH_2IBr}$) which we are familiar with\cite{gonzalez2010exploring, sandor2014strong,zhao2014removing}. One experiment focuses on differentiating photoelectron spectra according to ion momentum distribution, while the other on the momentum correlation in double ionization.

\section{Apparatus and Design}

\begin{figure*}
	\centering
	\includegraphics[width=\textwidth]{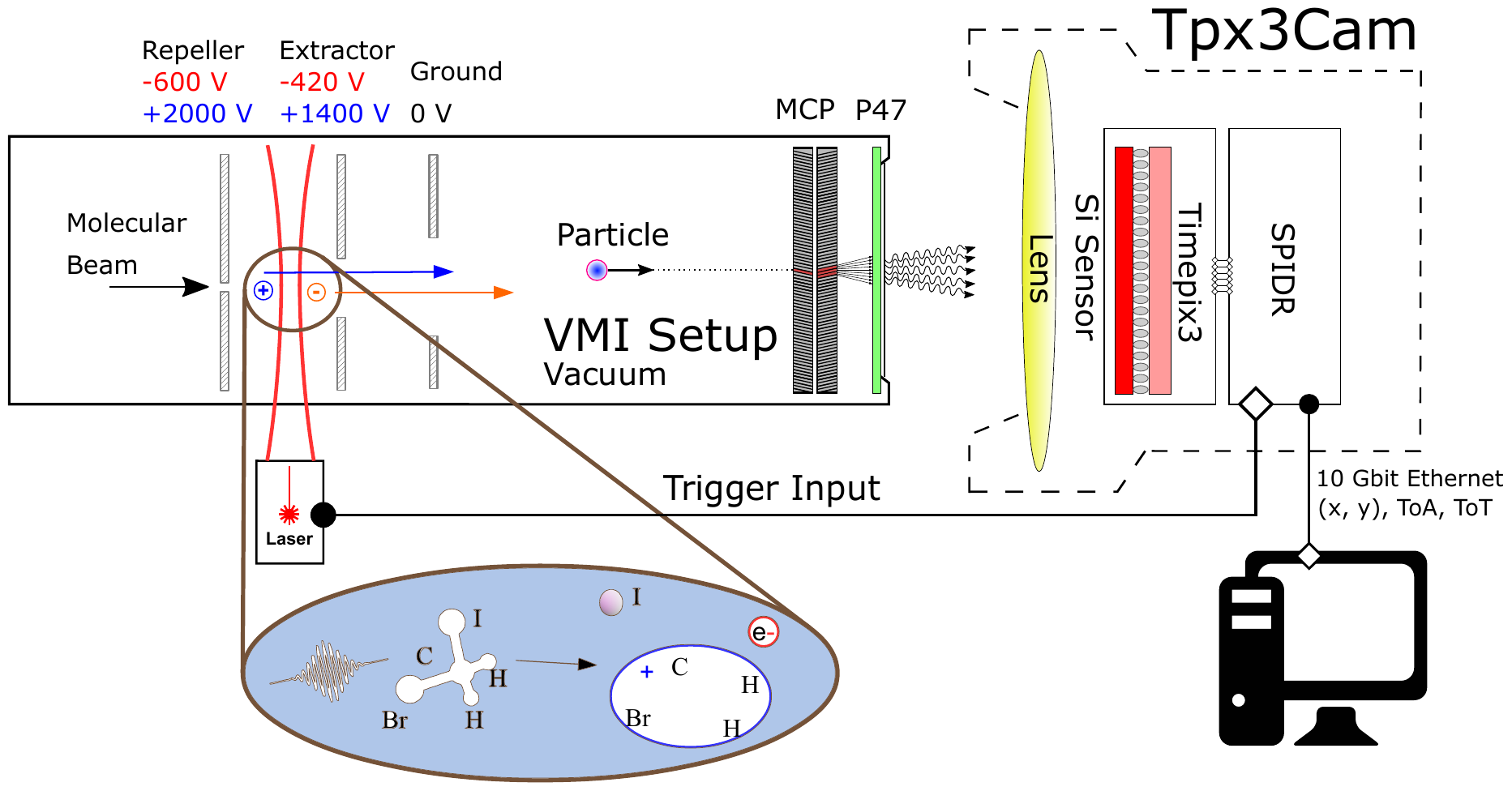}
 	\caption{Schematic diagram of the VMI apparatus with Tpx3Cam. The laser beam intersects an effusive molecular beam (eg. $\mathrm{CH_2IBr}$) the center of the VMI plates (Repeller, Extractor and Ground), producing photoelectrons (e-) and photoions ($\mathrm{CH_2Br+}$). The voltages on the VMI plates are switched from negative to positive immediately after the electrons leave the accelerating region. Due to large difference in mass, both electrons and ions are projected onto the dual stack of microchannel plates(MCP). Amplification in the MCPs leads to roughly 10$^6$ electrons hitting the P47 phosphor screen for every incident particle. The hit position (x,y) is imaged onto the Tpx3Cam sensor, which also records the time-of-arrival (ToA) and time-over-threshold (ToT). All information is uploaded to a computer for post-analysis.}
 	\label{fig:tpxcalibration}
\end{figure*}

A detailed description of the apparatus without Tpx3Cam, can be found in an earlier publication \cite{zhao2017coincidence}. Briefly, our light source starts with an amplified Ti:sapphire laser system, producing 30 fs (intensity FWHM) pulses with a central wavelength of 780 nm and a pulse energy of 1 mJ, at a 1 kHz repetition rate. The laser beam is then focused into an argon cell where new frequencies are generated due to self-phase modulation \cite{stibenz2006self}. The compressed pulses are measured to be sub-10 fs using self-diffraction frequency-resolved optical gating (SD-FROG). Inside a vacuum chamber, the laser intersects with an effusive molecular beam at the center of a set of VMI electrostatic plates (Fig. \ref{fig:tpxcalibration}). The voltages on the VMI plates are switched from negative to positive in less than 25 ns, in sync with the arrival of the laser pulse. Due to the large ion/electron mass ratio, the cations do not move much before the electrons leave the accelerating region and the voltages are then switched to direct the ions to the detector. This way, all charged particles are velocity mapped onto a dual stack of microchannel plates (MCP) in chevron configuration. A fast P47 phosphor screen behind the MCPs produces fast flashes of light for incident particles, which are imaged and time stamped by the sensor in the Tpx3Cam camera.

This design is relatively low cost, easy to set up, and allows data acquisition in both coincidence and non-coincidence modes\cite{zhao2017coincidence}. Since Tpx3Cam is placed outside of the vacuum, it is completely decoupled from the whole VMI setup, which brings considerable flexibility and allows for easy upgrades when improved cameras become available. Another advantage of this approach is that the camera optics allow one to map the whole phosphor screen on to a single sensor or to select arbitrary mapping arrangements employing appropriate lenses.

Tpx3Cam is a so-called hybrid pixel detector, which attaches a light sensitive, pixelized silicon sensor to the Timepix3 readout chip \cite{poikela2014timepix3}. Each pixel acts independently as a fast digitizer which can be used to record timing information. The new camera is capable of registering multiple hits with a time resolution of nanoseconds with a MHz rate, offering 10 times  better timing resolution and 100 times better throughput over TimepixCam\cite{fisherlevine2016timepixcam} that was previously used in time-stamping experiments \cite{hirvonen2017tcsp,chakaberia2017}.

\section{Tpx3Cam}

\begin{figure}
	\centering
	\includegraphics[width=0.90\linewidth]{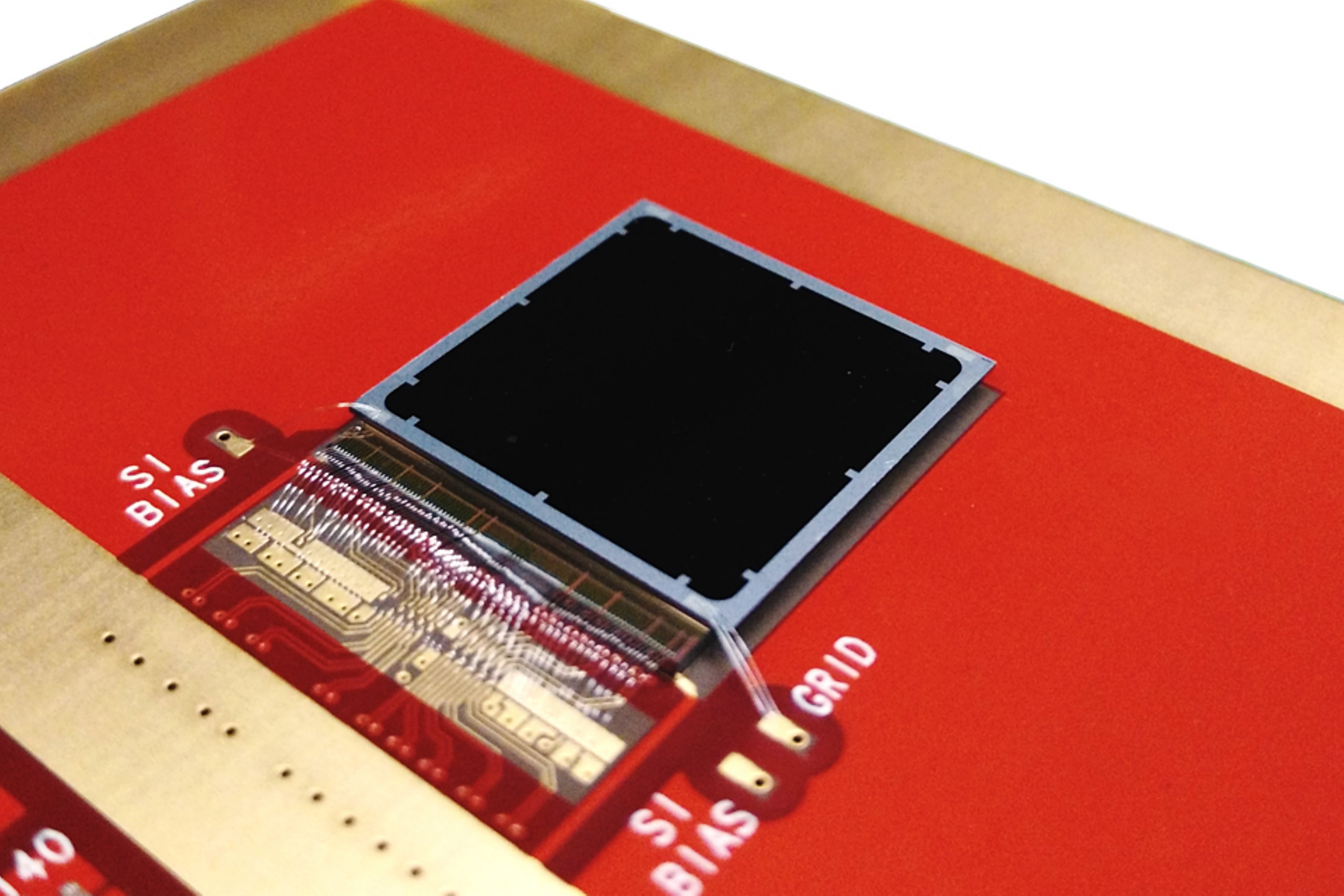}
	\caption{Photograph of the sensor assembly.}
	\label{fig:tpx3cam}
\end{figure}

The heart of the Tpx3Cam camera is a specialized silicon pixel sensor\cite{nomerotski2017characterization} bump-bonded to the Timepix3 readout chip. The sensor has thin entrance window with anti-reflective coating providing high quantum efficiency for light with wavelengths between 400 and 900 nm. For particle detection, the MCP is typically paired with a P47 phosphor, which has an emission maximum at 430 nm. Fig. \ref{fig:tpx3cam} shows a photograph of the sensor bump-bonded to Timepix3 on a chip board in the camera.

Timepix3 is a readout chip developed by the Medipix3 collaboration, which was used for the detection of X-rays and other types of ionizing radiation, but so far not for the optical photons. The signal in each pixel, after amplification, is compared to a predefined threshold. If the signal is higher than the threshold, its time-of-arrival (ToA), is measured together with the time-over-threshold (ToT). ToT provides an estimate of the  signal amplitude, or, in our case, of the number of photons absorbed by the silicon sensor in front of the pixel. Each pixel has a 4-bit digital-to-analog converter (DAC) to adjust the threshold so the response of all pixels is as uniform as possible. The average pixel noise is 60 electrons, and threshold dispersion after equalization is 35 electrons\cite{poikela2014timepix3}. The minimum number of photons  required to cross the threshold is approximately 600 photons at 430 nm but in practice the threshold is set to have a negligible dark count rate.

ToA information is recorded in a 14-bit register at 40 MHz and can be refined by additional 4 bits with a nominal resolution of 1.56 ns (640 MHz). The uniformity of time response is better than 2.1 ns over the whole imaging area. ToT is recorded in a 10-bit counter at 40 MHz rate.  The chip is designed in a 130 nm CMOS technology and contains $256 \times 256$ pixels, with each having dimensions of $55 \times 55~\mu$m$^{2}$. Tpx3Cam has a lens, which maps each pixel to a $0.18 \times 0.18$ mm$^2$ area of the phosphor screen in an optimized geometry.

The chip architecture  allows for sparse, data driven readout and can achieve a throughput of up to 80 Mpixels/sec. The readout is simultaneous with data acquisition, allowing each pixel to remain sensitive most of the time. The individual pixel deadtime is equal to the pixel ToT + 475 ns. The maximum rate of a single pixel is 625 kHz, while the maximum rate of a double column is about 1 MHz. This large throughput makes it important to mask all noisy pixels, otherwise they can  easily saturate the bandwidth, leading to large volumes of useless data. Timepix3 supports up to eight output data links with a total maximum bandwidth of 5.12 Gbps.

Tpx3Cam includes a general purpose readout system for pixel ASICs, SPIDR \cite{heijdenSPIDR}. The system is comprised of an FPGA board with memory and several communication interfaces including 1 and 10 Gigabit Ethernet. The latter allows a single  Timepix3 chip readout at the full bandwidth of 80 Mpixels/sec.

A data packet from a Timepix3 pixel has 48 bits containing both the ToT and ToA values. The ToA information is enough to bridge the hit transit time in the ASIC. In order to account for further transmission, the time-stamp information has to be extended. This happens in the  FPGA, which adds an additional 16 bits per pixel, such that each hit pixel produces a 64-bit entry.

SPIDR can also accept and time stamp an external trigger pulse, independent of the Timepix3 connection.  In our experiments, the trigger was synchronized with the firing of the laser shot, providing a time reference for registered electrons and  ions. The granularity of the trigger time measurement is 0.26 ns.

\section{Characterization of electron and ion detection}

\begin{figure}
	\centering
	\includegraphics[width=0.9\linewidth]{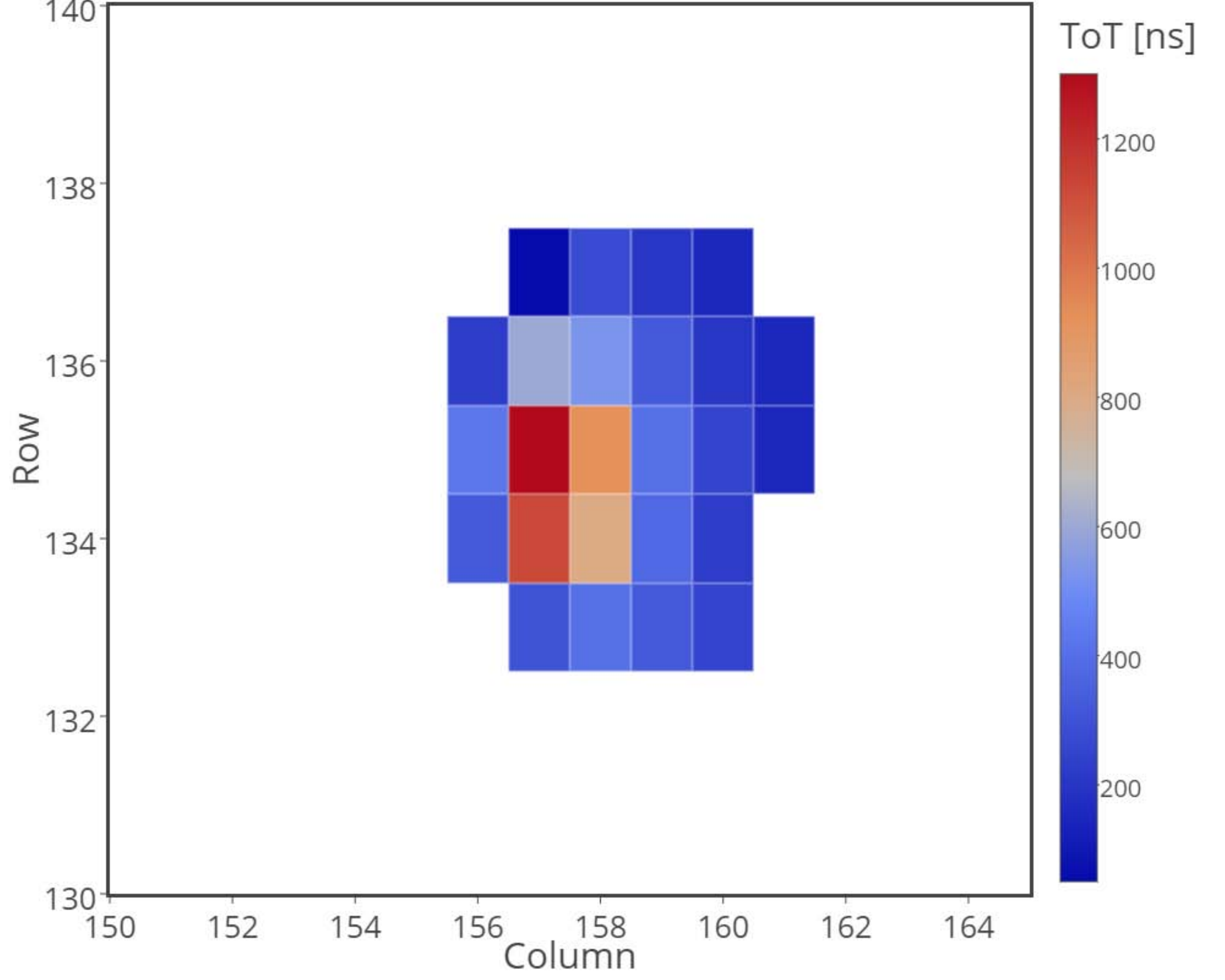}
    \includegraphics[width=0.9\linewidth]{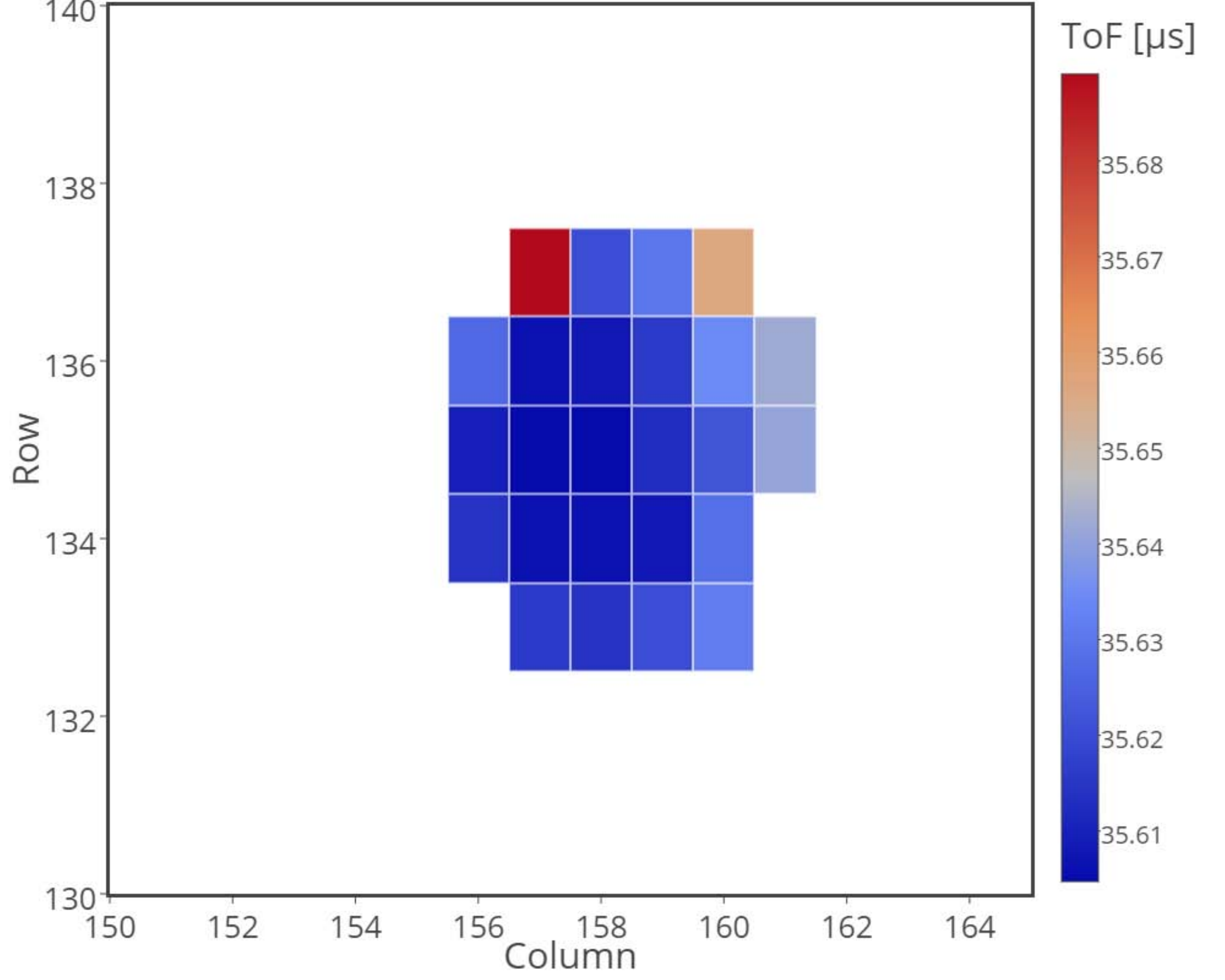}
	\caption{An example of ion hit in the camera in ToT (top) and ToF (bottom) color maps.}
	\label{fig:hits}
\end{figure}

Electrons and ions are registered by the camera as clusters of neighboring pixels, which all have signal above a threshold, with measured ToT and ToA. Fig. \ref{fig:hits} shows an example of an ion hit in the camera shown as ToA and ToT color maps. Typically the clusters have symmetric shapes with a few pixels having a large signal surrounded by pixels with a smaller signal. This is due to the symmetric nature of the charge multiplication in the MCPs, and the consequent photon emission in the phosphor imaged with the camera optics.

\begin{figure}
	\centering
	\includegraphics[width=0.98\linewidth]{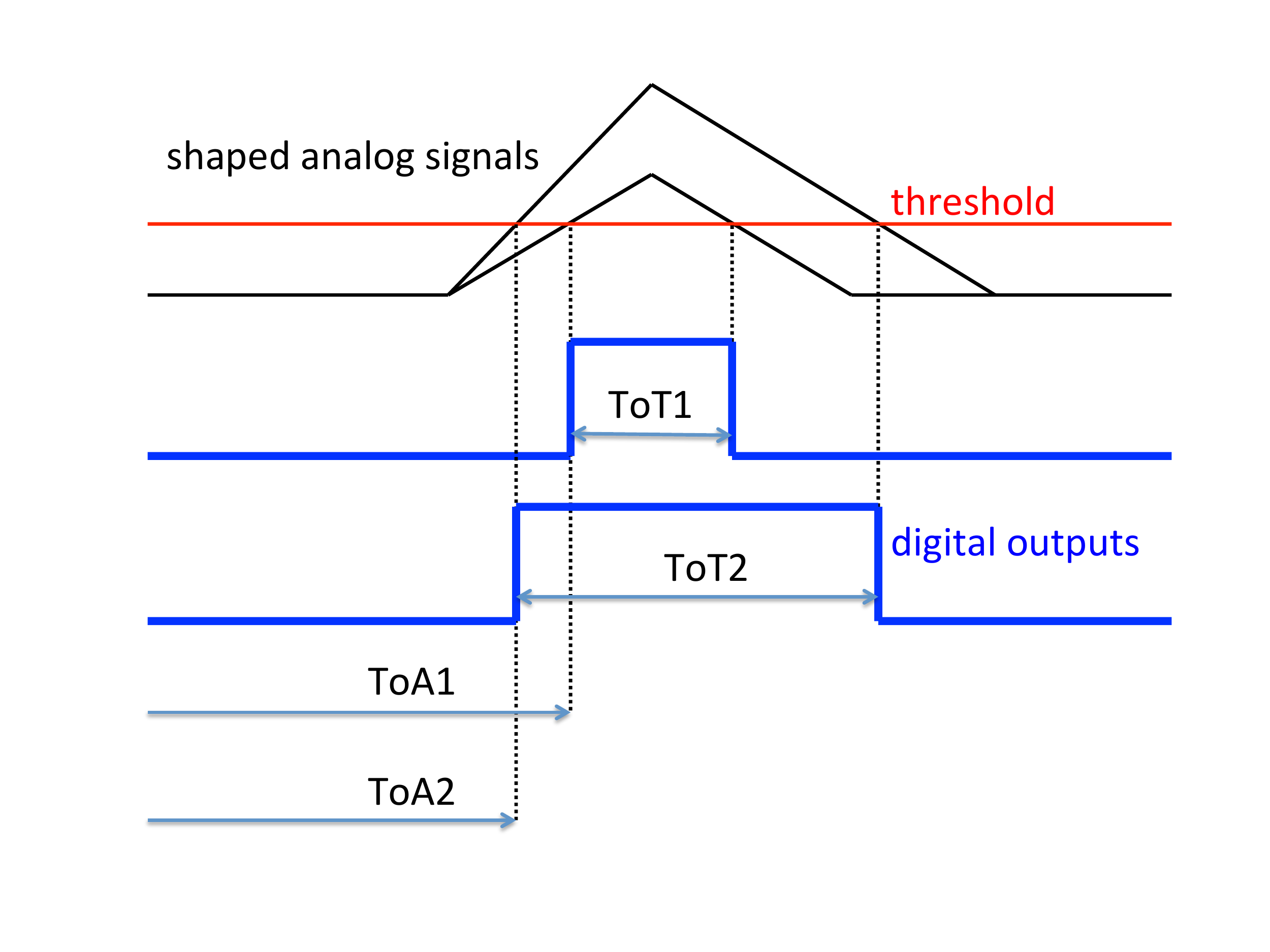}
	\caption{Time-walk in a discriminator with a constant threshold. Larger signals cross the threshold earlier, producing smaller ToA and larger ToT values.}
	\label{fig:CTD}
\end{figure}

\begin{figure}
	\centering
	\includegraphics[width=0.95\linewidth]{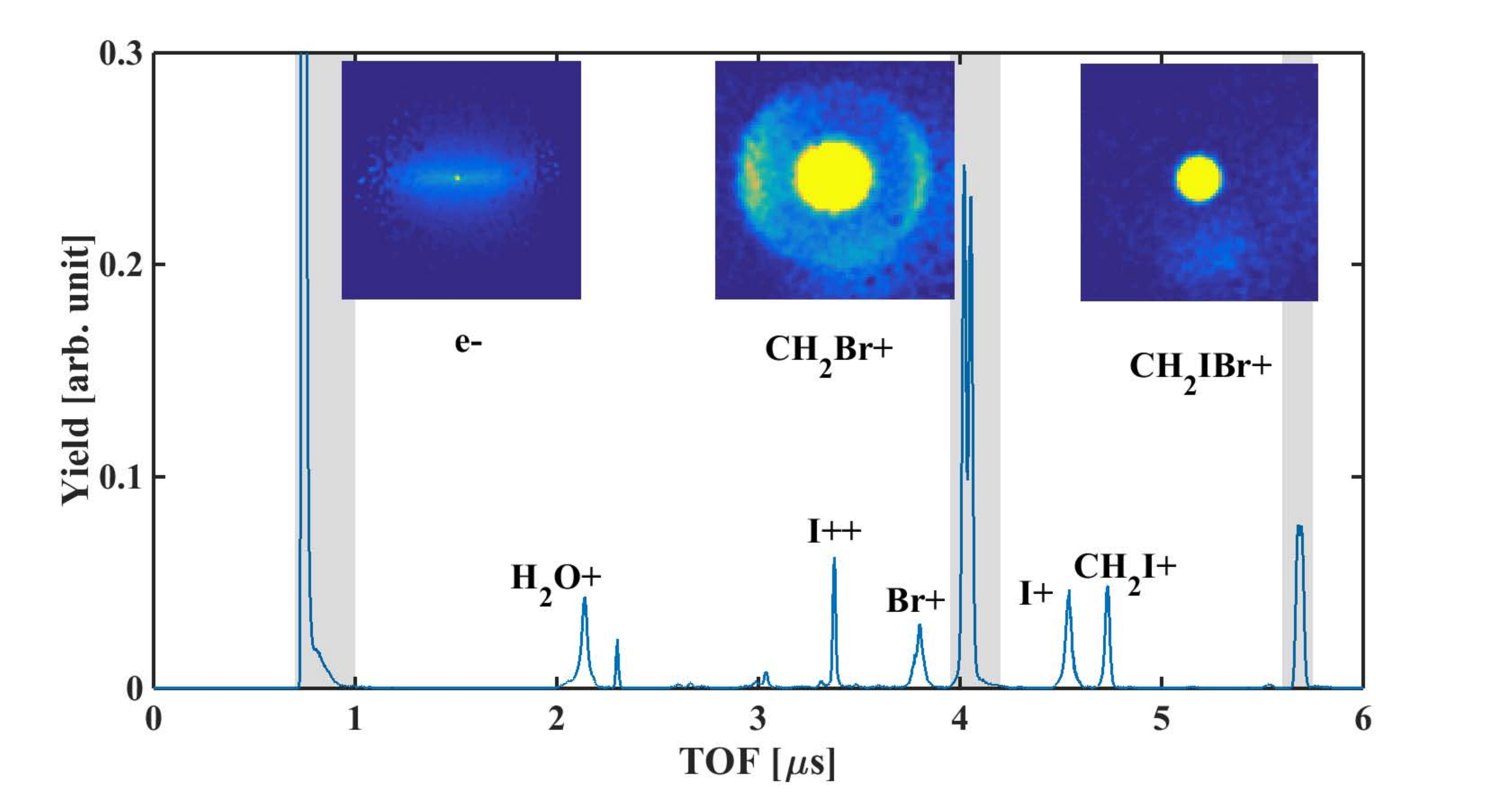}
 	\caption{ToF mass spectrum for CH$_2$IBr following strong field ionization. Momentum distributions are shown for selected fragments.}
 	\label{fig:TOF}
\end{figure}

It is apparent in the figure that pixels with the largest ToT correspond to pixels with earliest ToA. This is due to the so-called time-walk, a known effect in a constant threshold discriminator, illustrated in Fig. \ref{fig:CTD}. Signals on the output of the Timepix3 front-end electronics have similar rise time, and as a result larger signals cross the threshold earlier, producing smaller ToA and larger ToT values. Electrons and ions in the mass spectrum are identified by their time-of-flight (ToF), which is calculated as the difference between the trigger time and ion or electron ToA.

Each cluster of pixels is generated by a single particle with a certain ToF and coordinates on the MCP, so the information from multiple pixels can be used to improve the spatial and temporal resolution by applying a centroiding algorithm. First, a simple algorithm is employed to find clusters of connected pixels. Then the pixel with the highest ToT is found in each cluster, and the ToF of this pixel is taken as the particle's ToF. The position of the particle is also determined as coordinates of the pixel with the highest ToT.


The ToF spectrum from the Tpx3Cam data, with assignment of the mass peaks, is shown in Fig. \ref{fig:TOF}, together with the (x,y) distributions for electrons, CH$_{2}$Br$^{+}$, and CH$_{2}$IBr$^{+}$ mass peaks. The CH$_2$IBr molecule used for this strong field ionization experiment has multiple ion fragments corresponding to different ToF, and the figure demonstrates efficient registration of all its fragments by the camera in addition to the electrons. The centroiding algorithm described above was applied. Salient features of the graph are the ability to separate Br isotope peaks enabled by the excellent timing resolution of Tpx3Cam and variations in the momentum distributions for the fragments due to the different molecular dynamics leading to their formation.

\begin{figure*}[t]
  \centering
  \subcaptionbox{Electrons in raw data.}[.3\linewidth][c]{%
    \includegraphics[width=.32\linewidth]{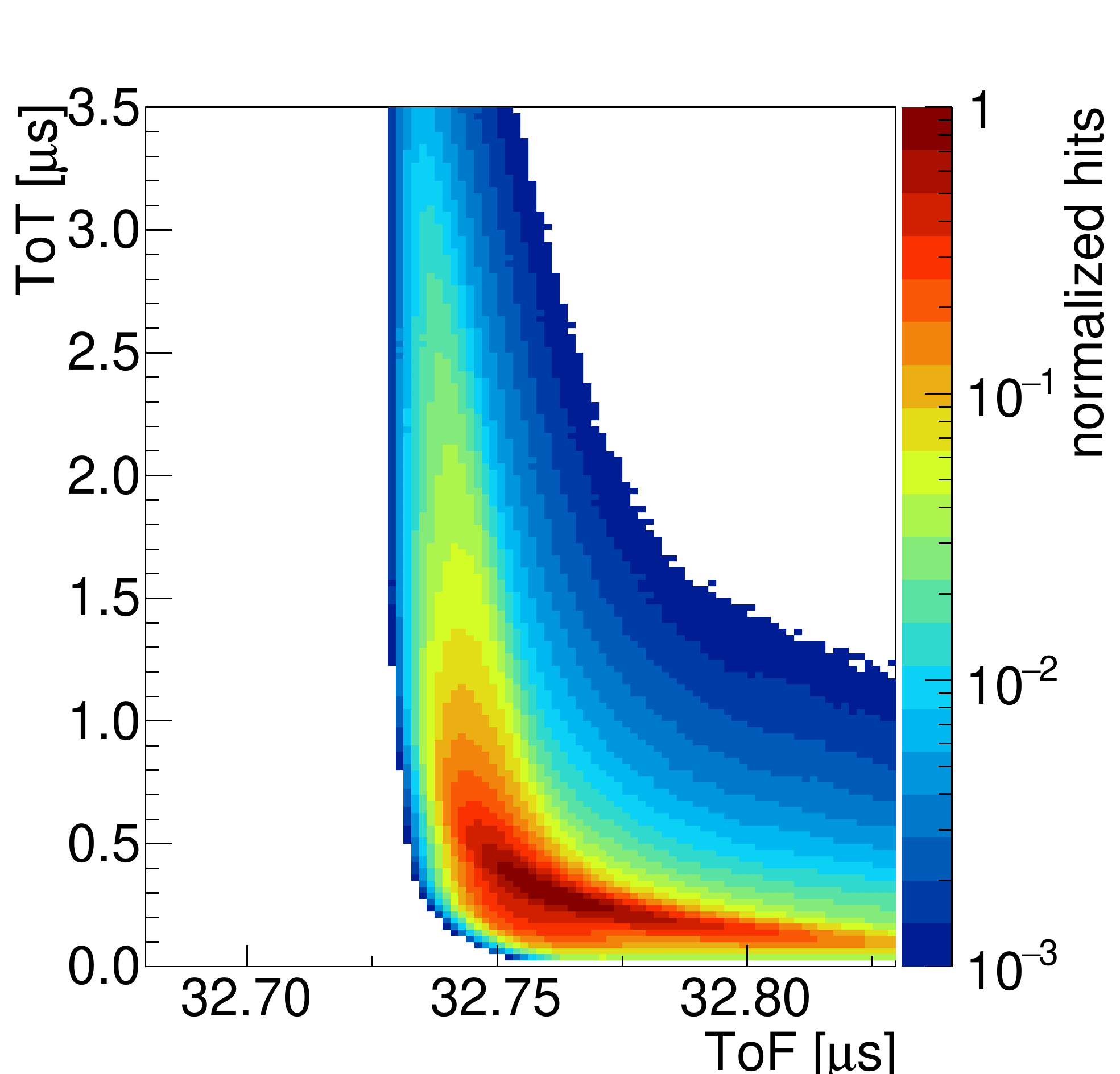}}\quad
  \subcaptionbox{Electrons after centroiding.}[.3\linewidth][c]{%
    \includegraphics[width=.32\linewidth]{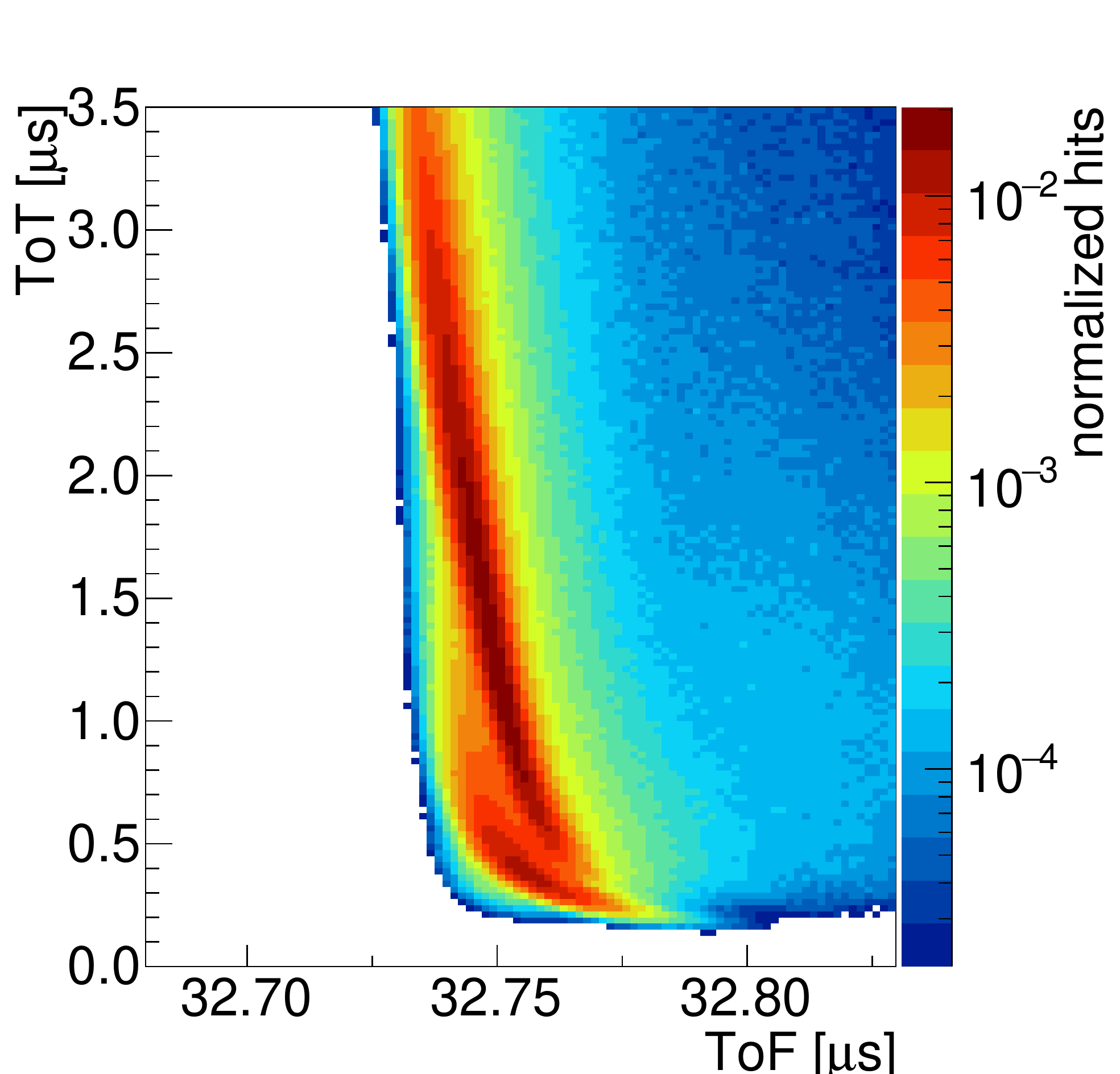}}\quad
  \subcaptionbox{Electrons after centroiding and ToT correction.}[.3\linewidth][c]{%
    \includegraphics[width=.32\linewidth]{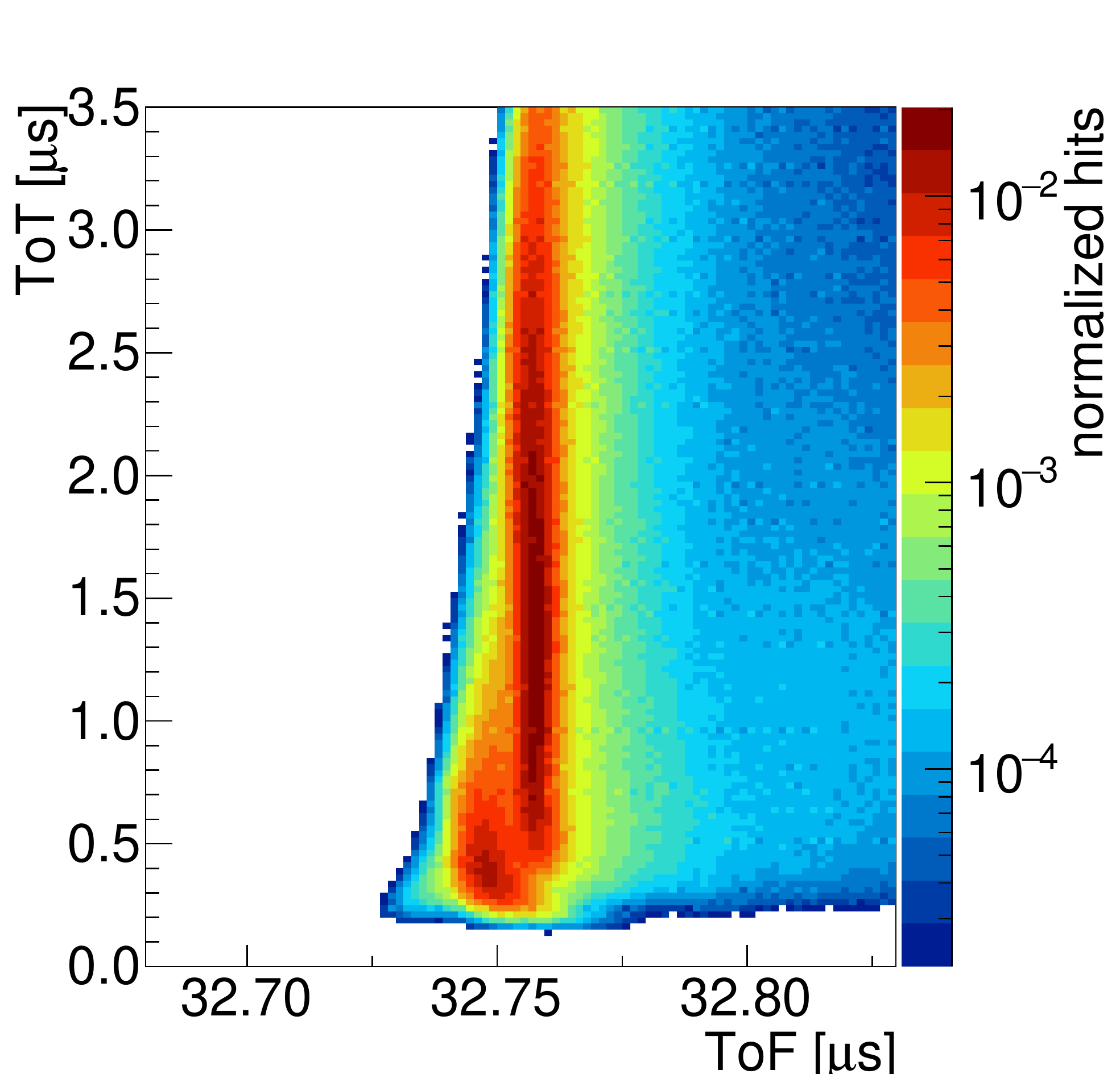}}

  \subcaptionbox{Ions in raw data.}[.3\linewidth][c]{%
    \includegraphics[width=.32\linewidth]{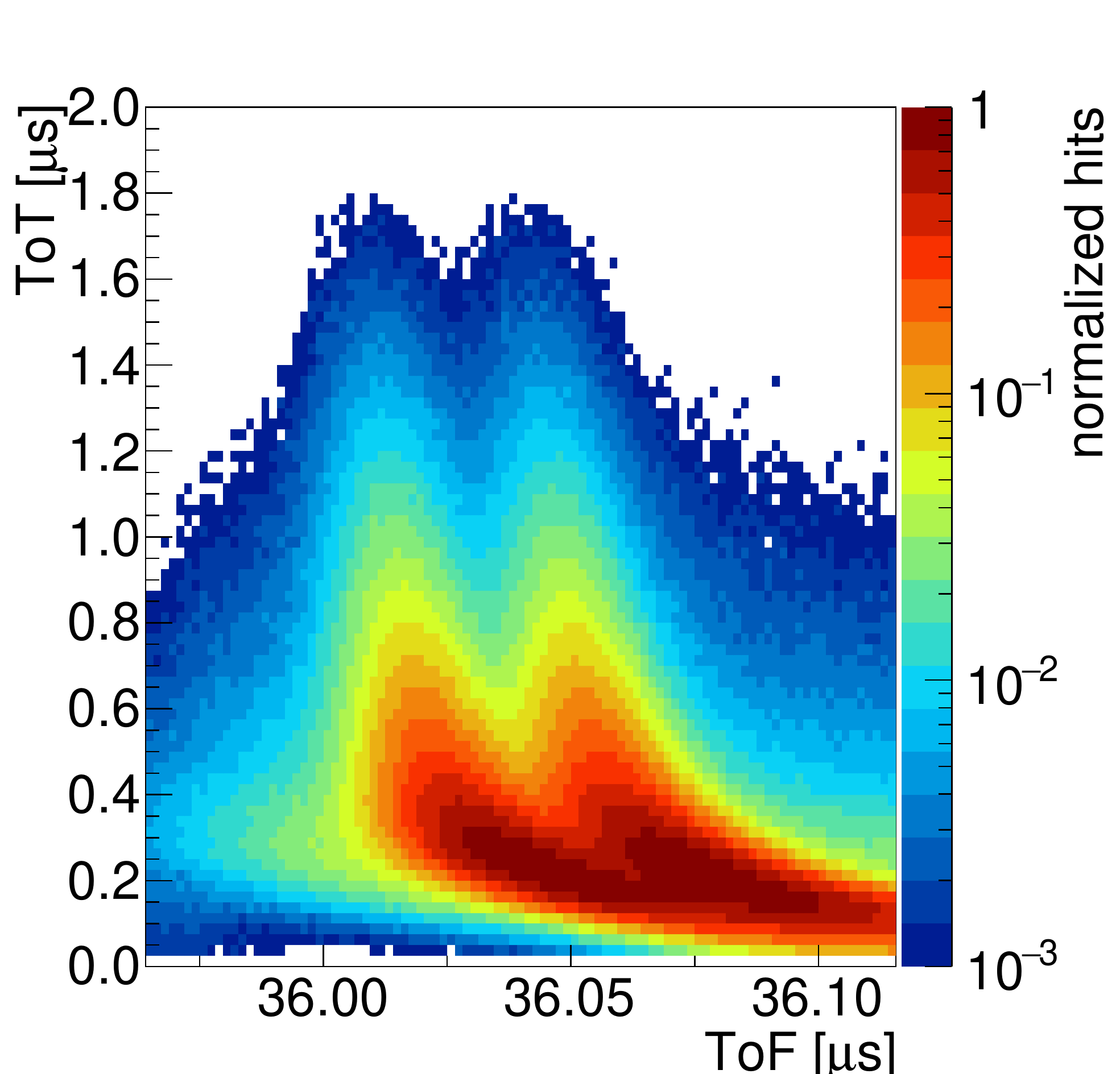}}\quad
  \subcaptionbox{Ions after centroiding.}[.3\linewidth][c]{%
    \includegraphics[width=.32\linewidth]{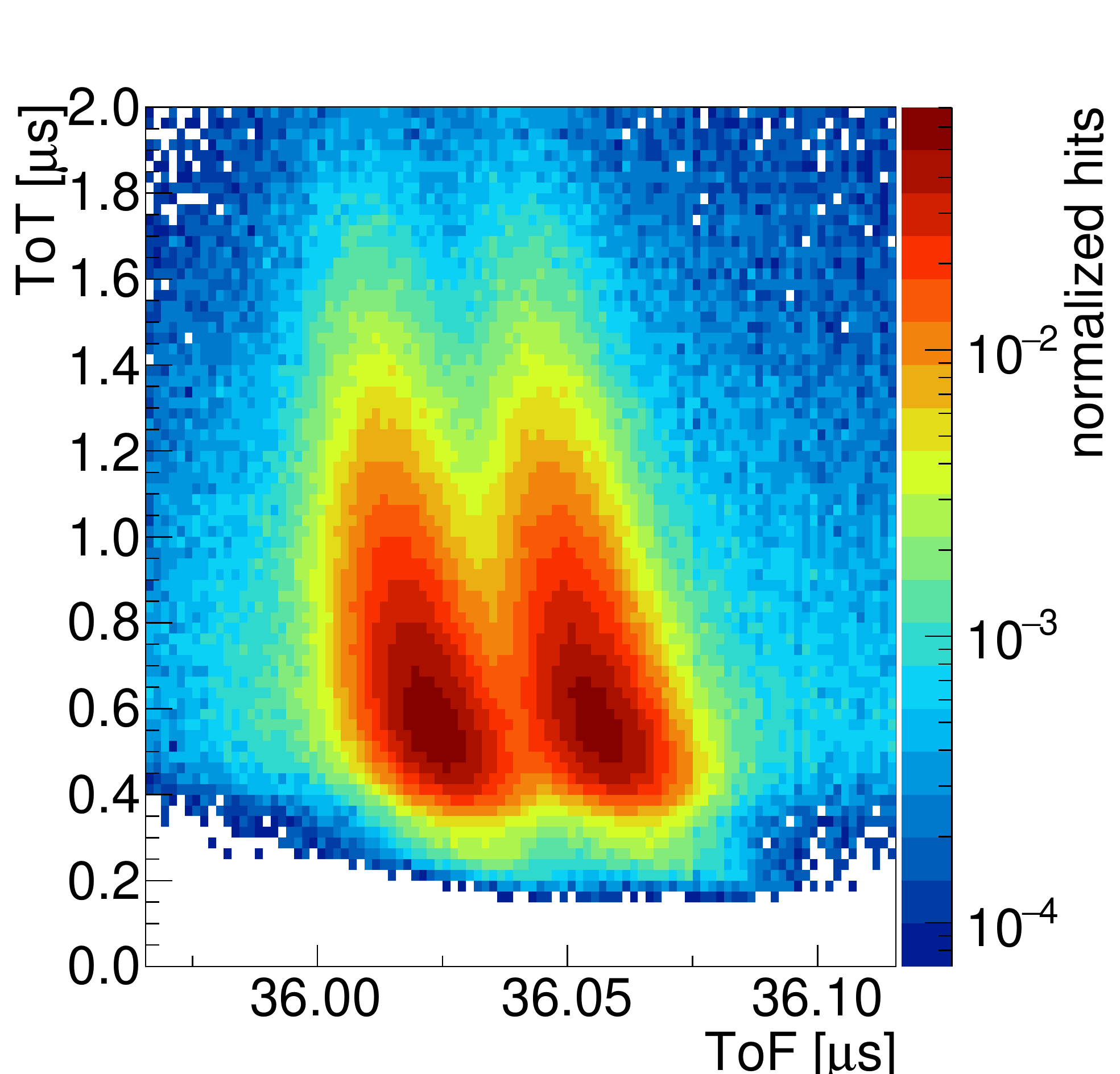}}\quad
  \subcaptionbox{Ions after centroiding and ToT correction.}[.3\linewidth][c]{%
    \includegraphics[width=.32\linewidth]{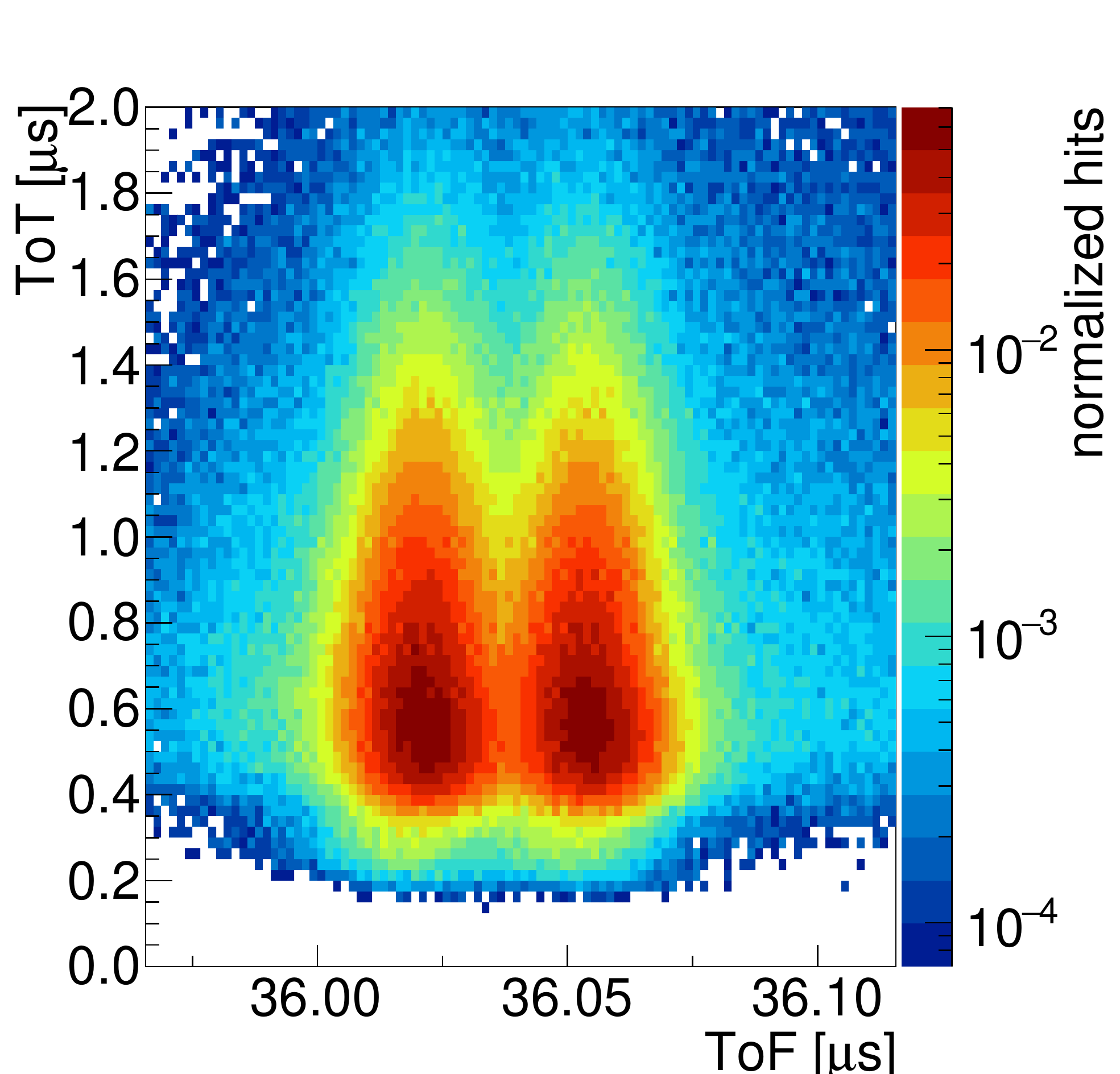}}
  \caption{ToT and ToF correlation for electrons and ions before and after centroiding and TOT correction.}
\label{fig:TOTTOF}
\end{figure*}

As explained above, larger signals result in faster ToF and longer ToT, so these quantities are correlated. Below we explore these correlations for different cases and explain how the correlations can be used to improve the time resolution of the camera. Fig. \ref{fig:TOTTOF} (a), (b), and (c) show  ToT and ToF correlations for the electron mass peak. Fig. \ref{fig:TOTTOF} (d), (e), and (f) show the correlations for the double mass peak of CH$_{2}$Br$^{+}$ corresponding to two Br isotopes, $^{79}$Br and $^{81}$Br. The correlations are shown for the raw data in (a) and (d), where all pixels were used to produce the graphs. The raw distributions are dominated by pixels with small ToT values, which have the largest time-walk. Fig. \ref{fig:TOTTOF} (b) and (e) show the same correlation after the centroiding algorithm was applied. One can see considerable drop in the population of small ToT values since the centroiding selects only  pixels with largest ToT in the cluster.

\begin{figure}
	\centering
	\includegraphics[width=0.95\linewidth]{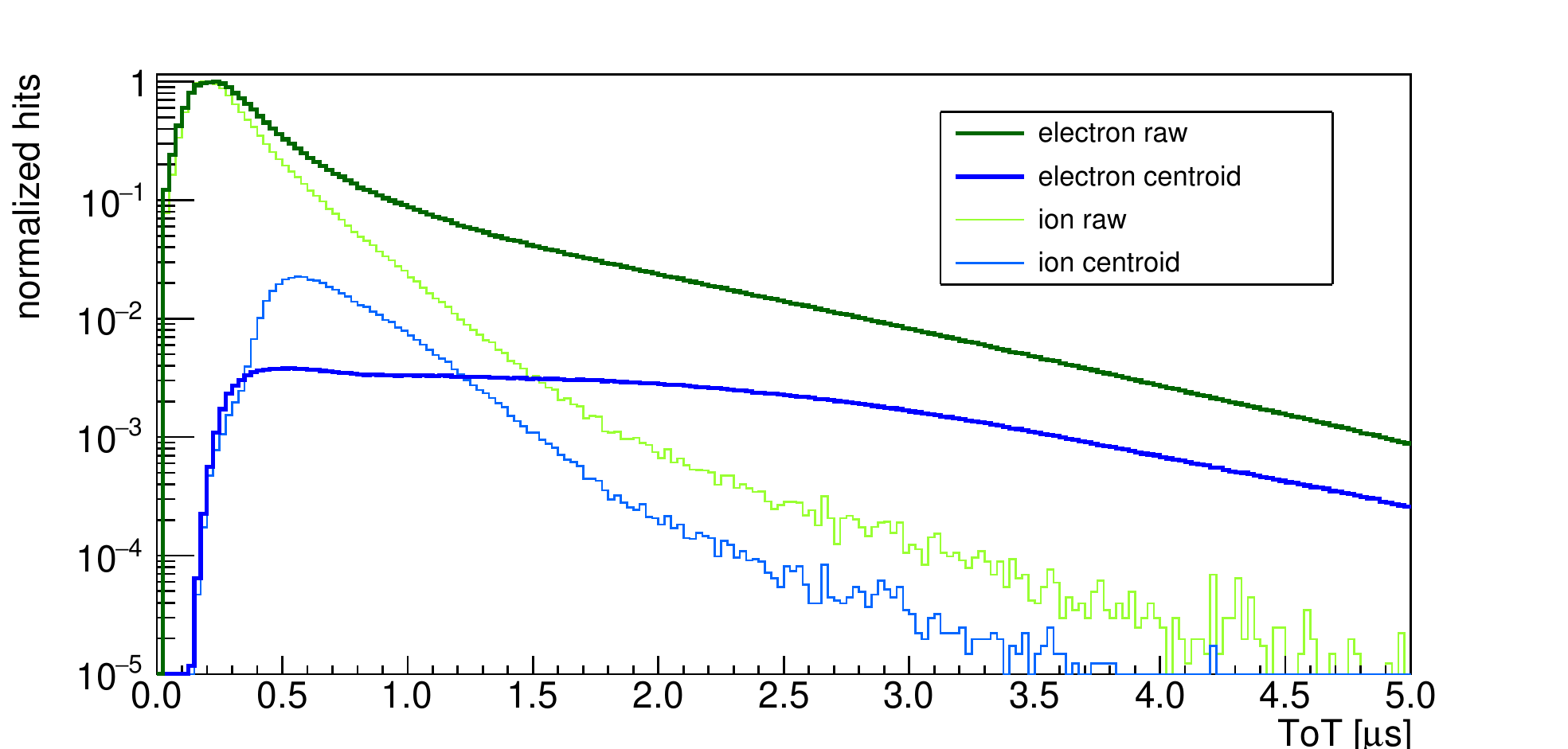}
 	\caption{ToT spectrum for electrons and I$^{++}$ ion before and after the centroiding.}
 	\label{fig:TOT}
\end{figure}

This is further illustrated in  Fig. \ref{fig:TOT}, which shows the ToT spectrum for electrons and I$^{++}$ ions before and after the centroiding. It is apparent that the ToT for electrons is considerably larger than that for the ions, presumably due to the surface saturation effects for the ions impinging the MCP, which results in a smaller number of primary electrons ejected from MCP.

In case of electrons, in Fig. \ref{fig:TOTTOF}(b), two bands become visible, one corresponding to the proper electrons and another one corresponding to the scattered laser photons, which precede the electrons by about 8 ns. Optical photons have very low efficiency and suppressed amplification in MCPs, leading to a lower amplitude signal compared to electrons. Nevertheless they can be clearly separated from the electron hits using the correlation plot.

The distributions discussed above show a clear dependence of ToF on ToT, which can be exploited to correct ToF and, therefore, to improve the time resolution. The correction procedure takes the electron peak and tabulates the average ToF corresponding to a narrow slice of ToT values. Then it calculates an offset needed to shift the pixel ToF in order to have the same ToF value for all ToT values. The results of this correction are shown in Figs. \ref{fig:TOTTOF} (c) and (f), for electrons and ions, respectively. The correction gives a vertical orientation to the electron, ion, and photon bands so their projection on the ToF axis has minimal spread.

\begin{figure}
	\centering
		\includegraphics[width=0.9\linewidth]{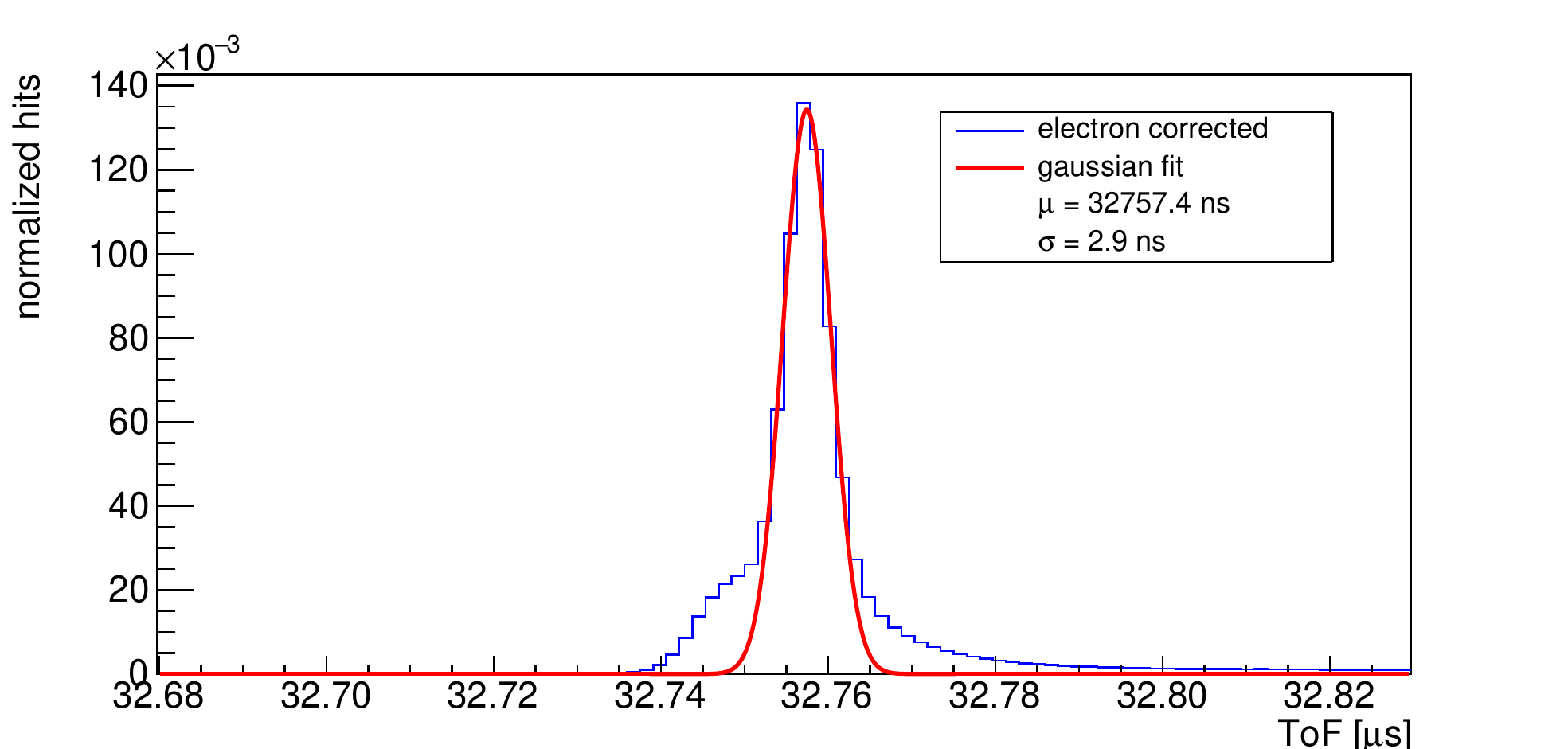}
    	\includegraphics[width=0.9\linewidth]{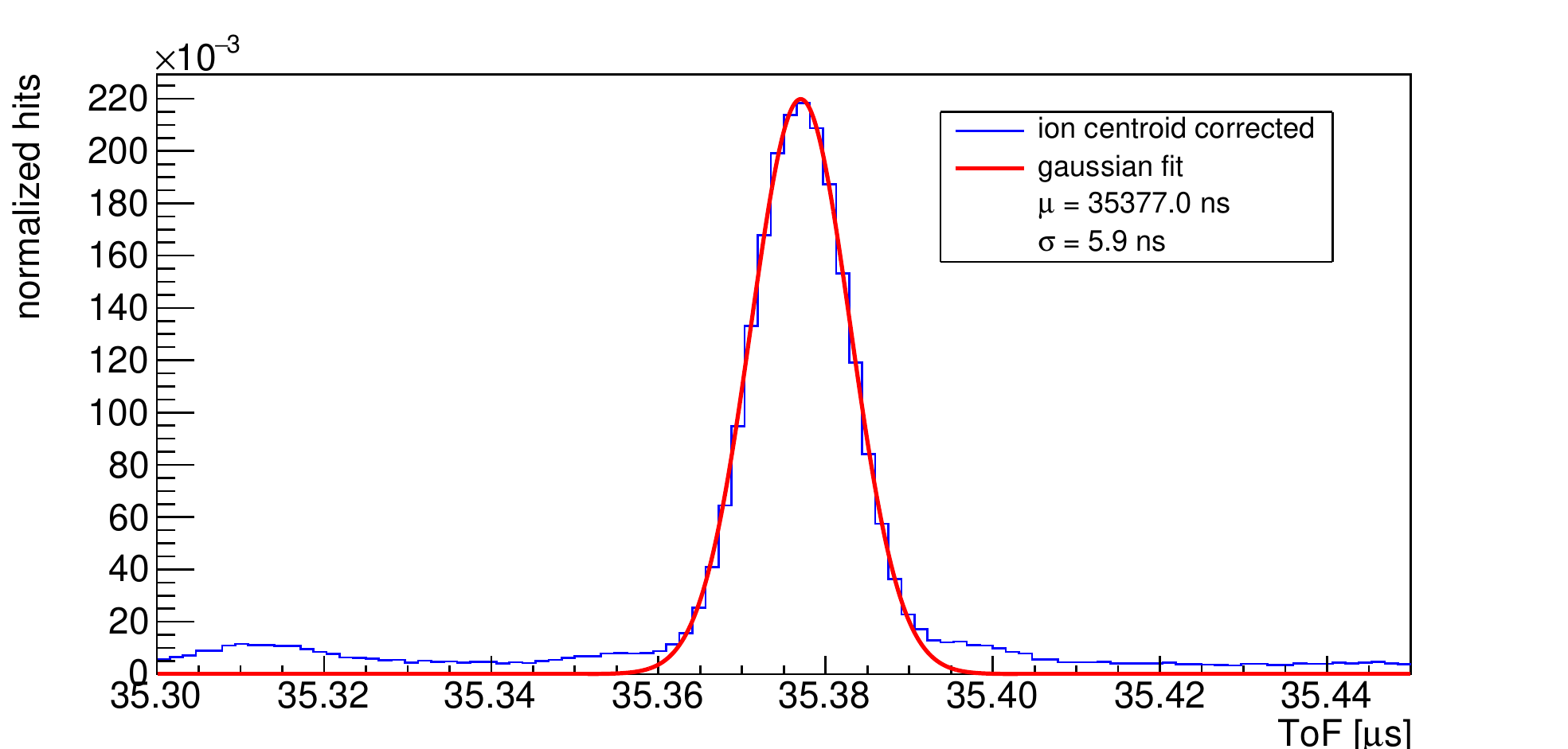}

\caption{Time resolution for electrons (top) and ions (bottom), both after centroiding and ToT correction.}
	\label{fig:timeresolution}
\end{figure}

The time resolution achieved after the ToT correction is given in Fig. \ref{fig:timeresolution} for the electrons and for the I$^{++}$ ion peak. The shoulder on the left of the electron mass peak is explained by the scattered photon signal, which is not so apparent in the projection as in the 2D correlation plot in Figs. \ref{fig:TOTTOF} (b) and (c). The  peaks are fit to Gaussian functions with sigmas of 2.9 ns and 5.9 ns for the electrons and ions respectively. The time resolution for electrons is considerably better than that for the ions due to larger signal produced by the electrons, which suppresses the time-walk related time jitter. The ions also have additional velocity spread upon their creation, which lead to additional time jitter. Without the ToT correction the time resolution is about 10 ns for both electrons and ions.

Note that the measured time resolution has contributions not only from Timepix3 but also from  variations of ToF inside the VMI apparatus and finite rise time of the P47 phosphor signal, which is about 7 ns\cite{winter2014p47}. The Timepix3 time bin, 1.56 ns, which corresponds to a standard deviation of 0.45 ns, should not be a limiting factor so we conclude that the time resolution in this setup is determined by the other factors mentioned above and, possibly, by the remaining residuals of the time-walk ToT correction.

\begin{figure}
	\centering
		\includegraphics[width=0.9\linewidth]{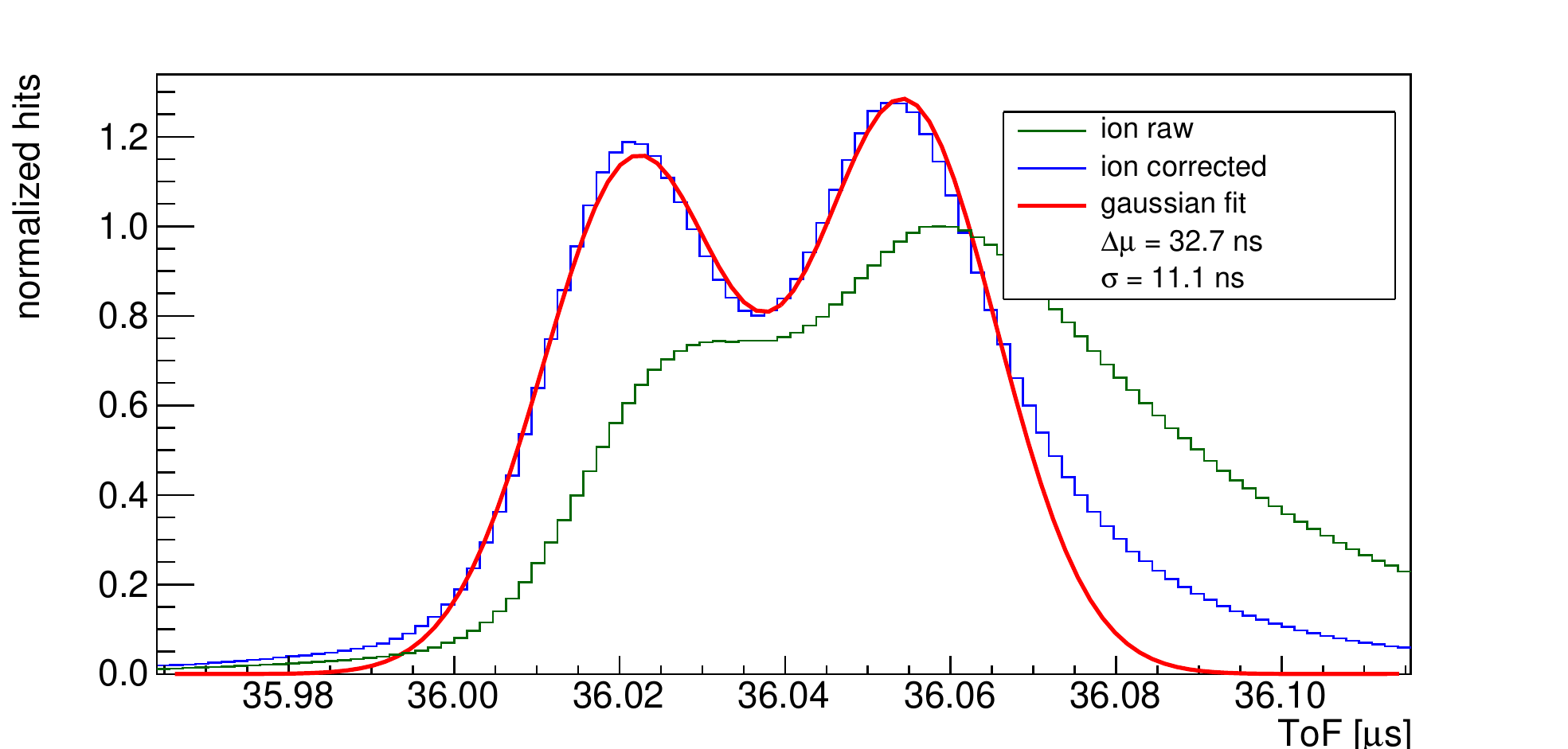}
    	\includegraphics[width=0.9\linewidth]{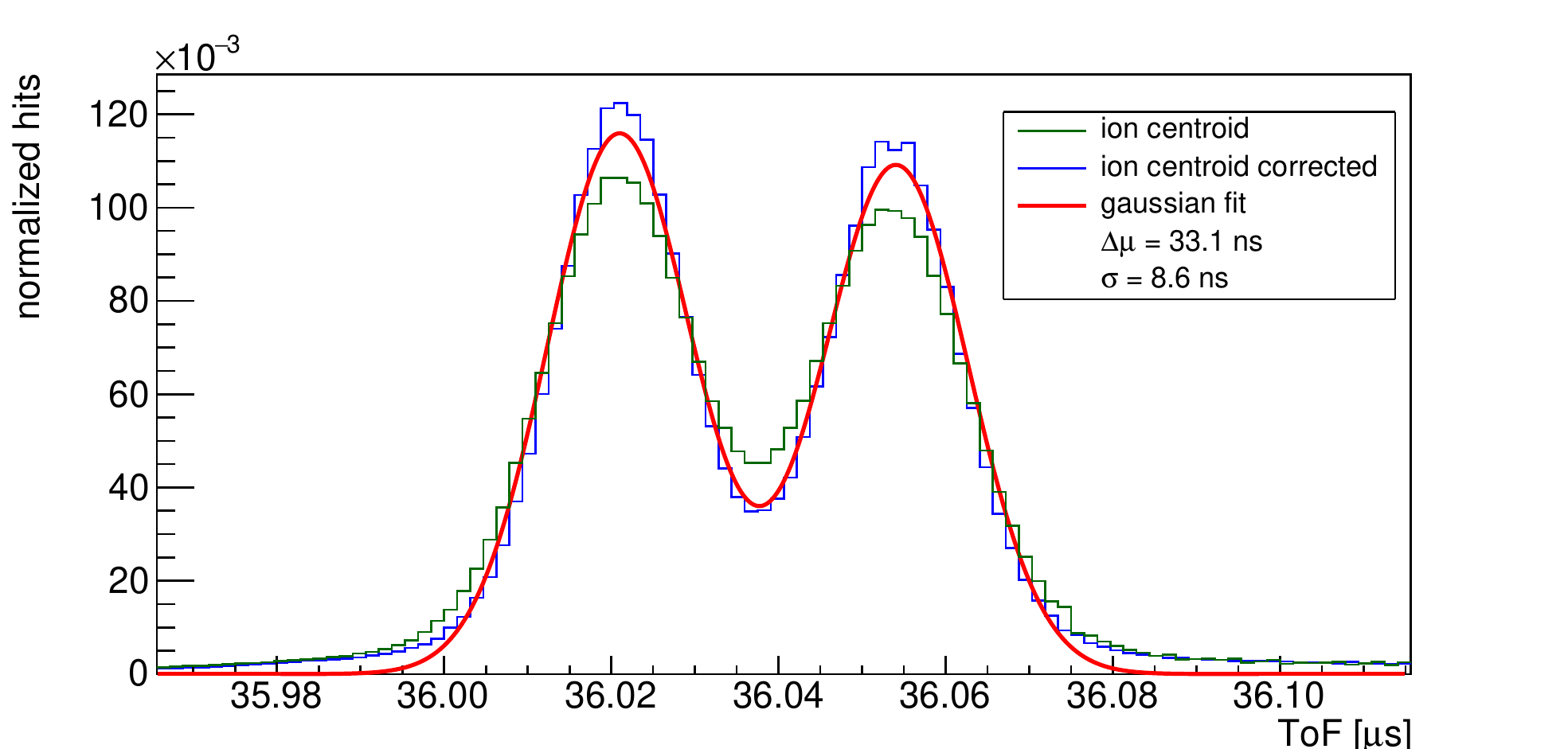}
	\caption{Mass peaks of two Br isotopes for raw pixel data (top) and centroided data (bottom) shown before (green line) and after (blue line) ToT correction. Corrected data is fit with a double Gaussian (red line).}
	\label{fig:timeresolution_double}
\end{figure}

Fig. \ref{fig:timeresolution_double} shows the time resolution after the ToT correction for the double mass peaks of the two Br isotopes. A reshaping of the peaks from the raw data histogram, that includes all pixels in the centroiding, to ToT corrected histogram is evident. It is clear that the centroiding which selects the largest ToT pixel has a dramatic effect on the time resolution, and ToT correction further improves it. The measured ToF difference of 33.1 ns between the two isotopes, $^{79}$Br and $^{81}$Br, and the 8.6 ns time resolution allow for an estimation of the resolving power $M/\Delta M = 154$.

\section{Results}

In this section, we present two measurements carried out with this new apparatus. In the first experiment, we measure the photoelectrons in coincidence with photoions resulting from the strong field single ionization of the bromoiodomethane molecule, $\mathrm{CH_2IBr}$, which has been used extensively in similar experiments and whose electronic structure is well studied \cite{sandor2014strong,gonzalez2010exploring}. Therefore this experiment serves both as a test of the new apparatus and a means of imaging calibration. The results are consistent with previous measurements, and additionally exhibit new features which we have not been able to observe previously with a conventional CMOS camera.

\begin{figure}[h!]
	\centering
	\includegraphics[width=0.8\linewidth]{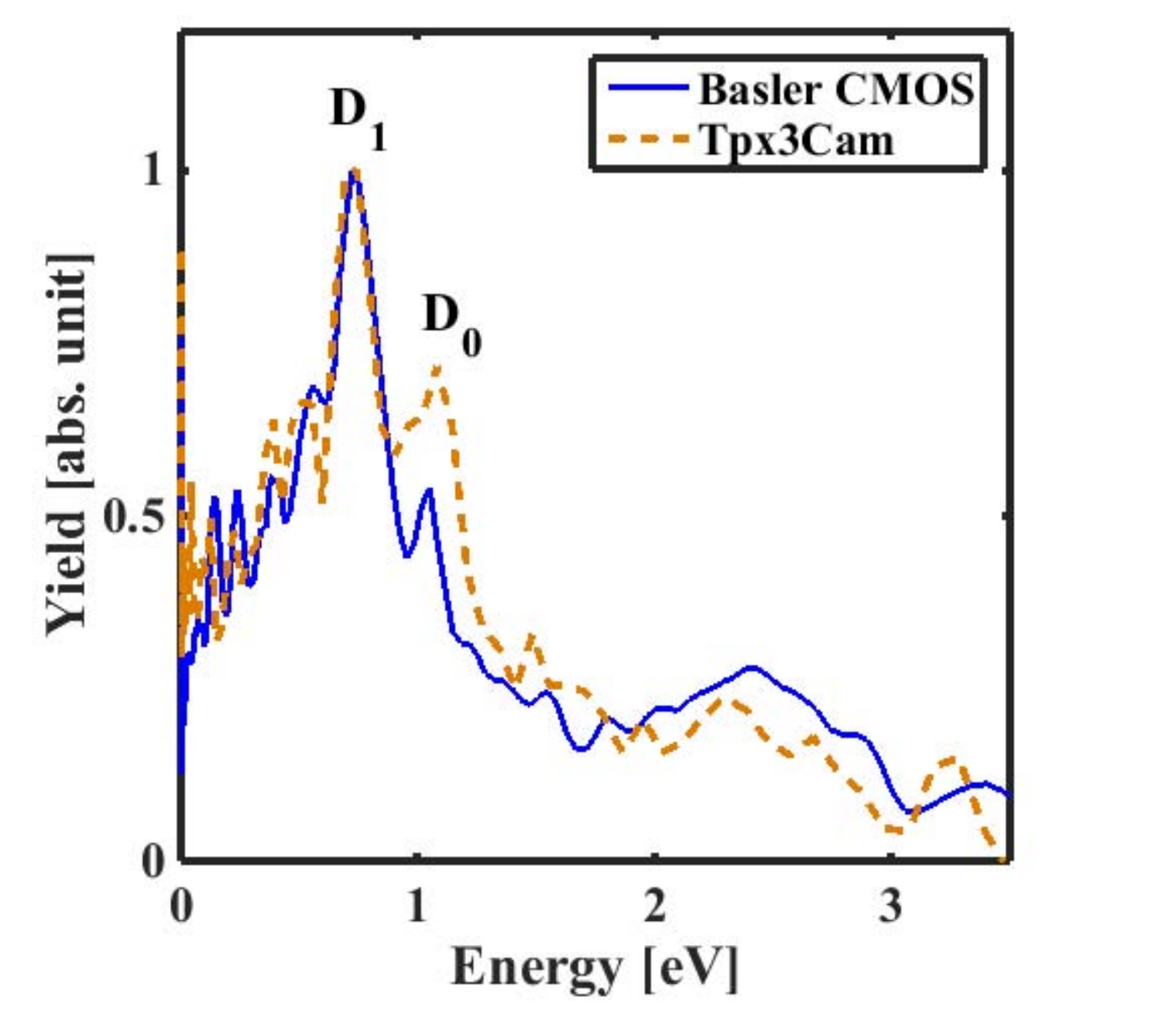}
	\caption{Comparison of photoelectron spectra measured with traditional CMOS camera (Basler acA2000) and Tpx3Cam camera. These are not coincidence measurements and the spectra contains photoelectrons from all ionization events. We see very similar structures in the spectra and these features provide energy calibrations -- mapping pixels to energy. The original energy calibration using the CMOS camera utilizes above-threshold ionization (ATI) peaks \cite{sandor2014strong}.}
	\label{fig:tpxcomparison}
\end{figure}

In Fig. \ref{fig:tpxcomparison}, we compare the photoelectron spectra obtained with a conventional fast CMOS camera (Basler acA2000) and the Tpx3Cam camera. The electrons are collected in non-coincidence mode and the raw images are Abel-inverted to recover the 3D momentum distribution. Angular integration for a constant radius renders the spectra shown in the Fig. \ref{fig:tpxcomparison}. Based on earlier experiments\cite{sandor2014strong}, we expect two peaks at around 1 eV, which are labeled $\mathrm{D_0}$ and $\mathrm{D_1}$ in the figure, denoting photoelectrons resulting from ionization to the ground and first excited ionic states, respectively. The similarity between these two spectra indicates that the new design performs well. Small discrepancies are believed to be a result of intensity variation and the noise introduced by the Abel inversion.

\begin{figure*}[h!]
	\centering
	\includegraphics[width=1\linewidth]{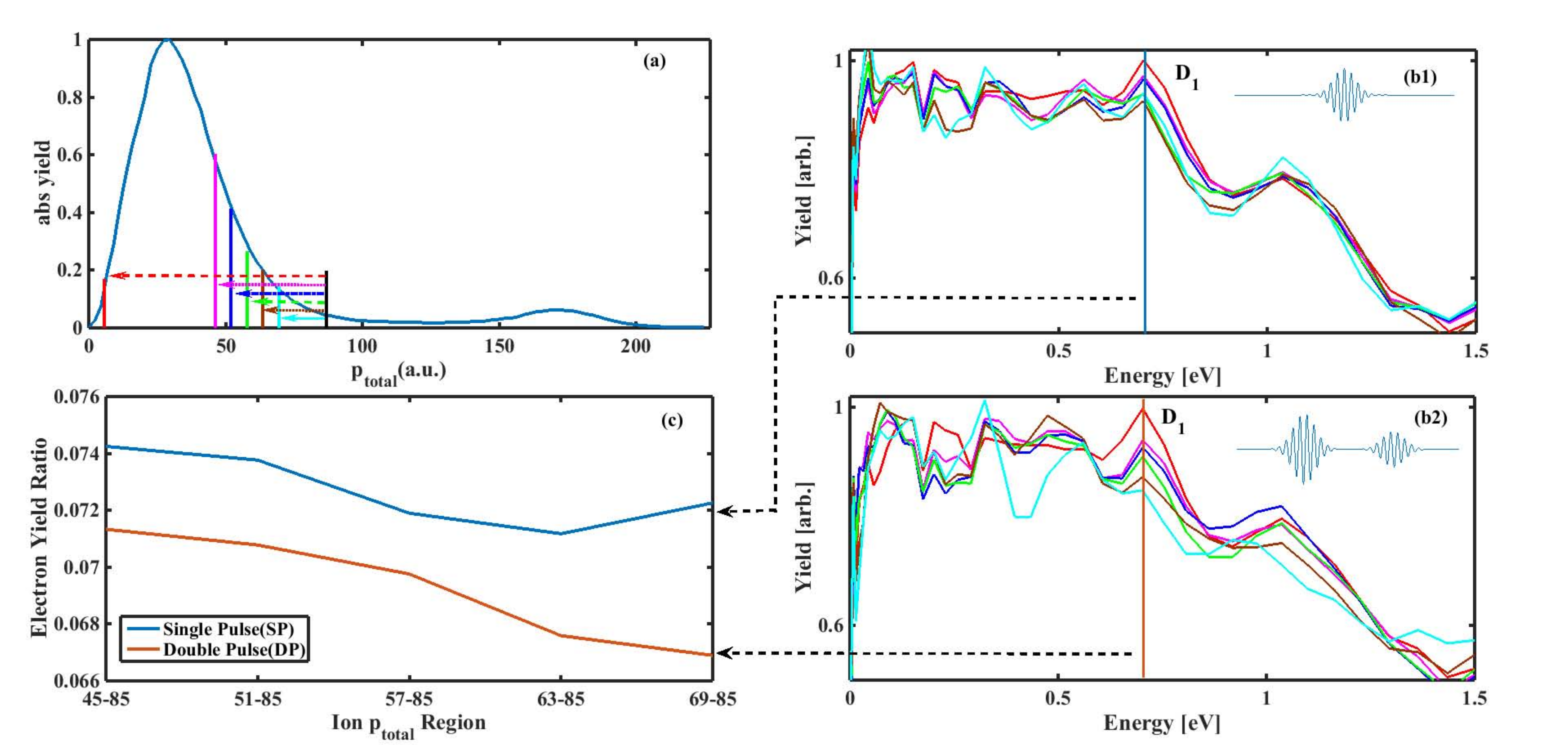}
	\caption{Illustration of electron spectra correlated with fragment ion momentum.   Photoelectron spectra correlated with different ionic fragment momenta are shown for different pulse shapes - single and double pulses. (a) Momentum distribution for  $\mathrm{CH_2Br^+}$ ions, along with indications of how events were selected to generate photoelectron spectra coincident with ions within a selected momentum range.  We first separate the ions into different momentum groups, corresponding to integration from high momentum (just under 100 a.u) to lower values indicated by the vertical lines. We then plot the photoelectron spectra measured in coincidence with ions in each of these region, using a single pulse (b1) and a `pump-probe' double-pulse (b2). (c) Integration of electrons in the D$_1$ peak divided by the total yield as a function of ion momentum.  Electrons generated with a single pulse show different behavior with the lower limit on ion integration than electrons generated with a double pulse.}
	\label{fig:tpxsi}
\end{figure*}

Now we present a measurement which demonstrates the ability of coincidence detection, facilitated by the Tpx3Cam, to make a highly differential measurement which isolates electrons in coincidence with fragment ions having specific momenta.  This allows one to see pulse shape dependent features in the ionization yield which would otherwise not be discernible. We choose a pump-probe pulse pair, for which we expect dynamics in the ground state of the molecular cation to produce fragment ions and photoelectrons whose momenta are correlated \cite{gonzalez2010exploring}.   The main result is shown in Fig. \ref{fig:tpxsi}. Here we compare photoelectrons in coincidence with $\mathrm{CH_2Br}^{+}$ (the most abundant fragment ion),  produced by a single pulse (SP) and by a pump-probe double-pulse (DP). In the DP experiment, the first pulse (pump) is identical to that used in the SP experiment, while the second pulse (probe) is 10 times less intense than the pump. The pump pulse ionizes the molecule and the probe redistributes population between ionic states, producing correlated changes in the electron and ion momentum distributions. The SP experiment is used as a reference for comparison.

We start with the 2D raw images of $\mathrm{CH_2Br}^{+}$, which show little difference for SP (right half of panel (a)) and DP (not shown) experiments. We select a series of concentric annular regions, with fixed outer radius (15 pixels) and decreasing inner radius, illustrated by a series of annular arcs color-coded by their inner radii on the left half of panel (a). Then we plot the spectra of the photoelectrons in coincidence with the ions that fall into these annular regions, panel (b1) for SP and panel (b2) for DP. Note that the spectrum is normalized by the number of electrons, producing what can be thought of as the spectrum for a single electron. As we restrict the ions to increasingly larger momentum ranges (from magenta to blue, green, brown and cyan), the $\mathrm{D_1}$ peak around 0.7 eV is increasingly suppressed in the DP experiment while remaining mostly constant in the SP experiment. In contrast, there is little change in the $\mathrm{D_0}$ peaks around 1.1 eV for both SP and DP experiments. Plotting the $\mathrm{D_1}$ peak values from both experiments in panel (c), we see a clear difference in the yields. Note that a significant change in the $\mathrm{D_1}$ yield for DP appears at region IV, 10 - 15 pixels away from the center. While the interpretation of the correlated changes in the electron and ion momenta with the application of a double pulse vs a single pulse are beyond the scope of this manuscript, we note that they are only detectable with a coincidence velocity map imaging apparatus that measures both electrons and ions with good VMI resolution.




\begin{figure}[h!]
	\centering
	\includegraphics[width=0.9\linewidth]{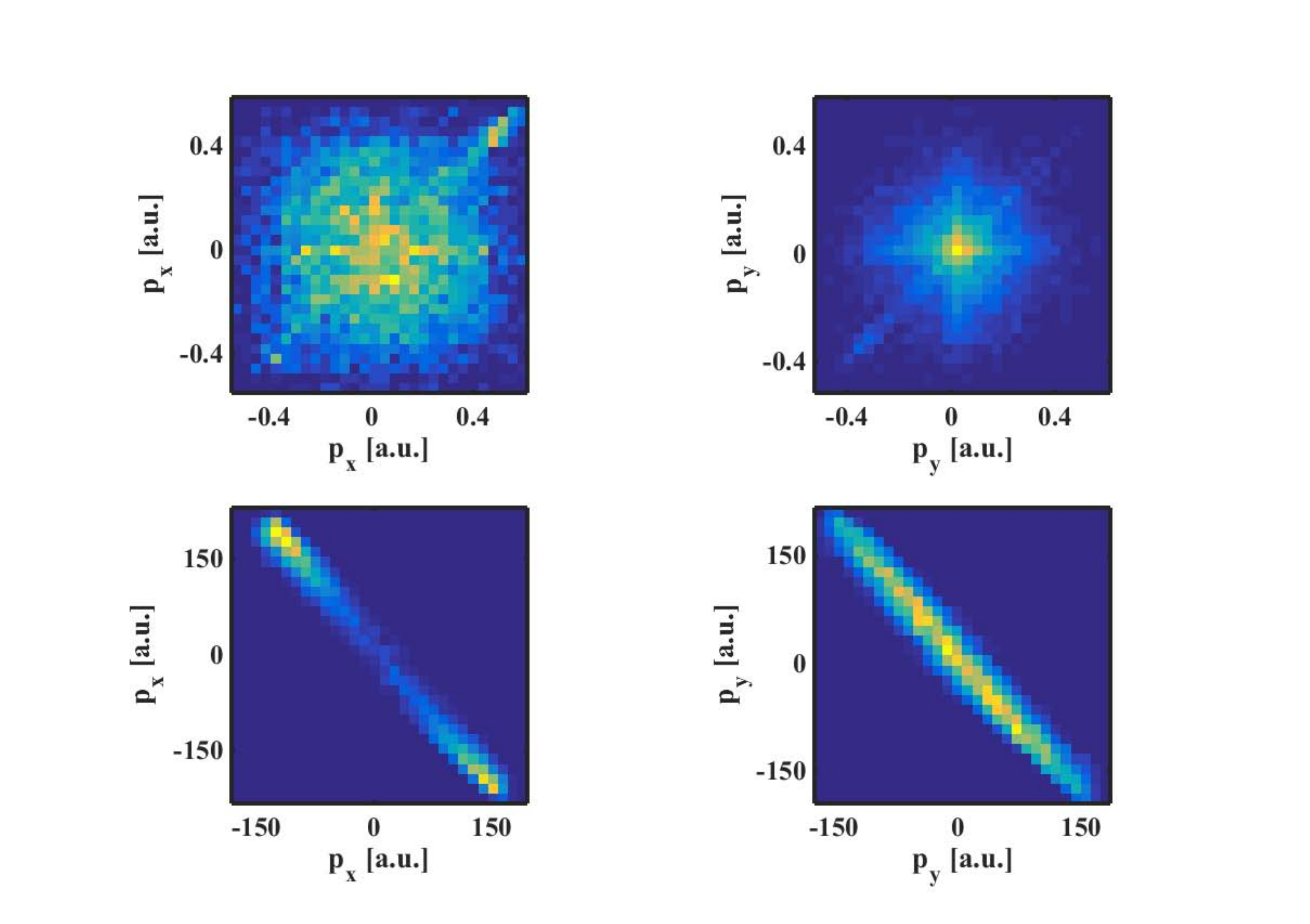}
	\caption{Electron-electron (top panels) and ion-ion (bottom panels) momentum correlations. The x-axis is chosen to be along the laser polarization direction. Ion-ion anti-correlations are expected due to momentum conservation while positive e-e correlations indicate two electrons leave the molecule together in the same direction.}
	\label{fig:tpxdi}
\end{figure}

A second experiment focuses on double ionization (DI) in $\mathrm{CH_2IBr}$. In fact, the DI results shown in Fig. \ref{fig:tpxdi} and the single ionization (SI) presented earlier come from the same data set. At the intensity ($\sim$ 10 TW/cm$^2$) where the measurement is carried out, about 5\% of laser shots generate a coincidence event. Most of these events are SI, containing exactly one electron and one ion. A DI event should produce exactly two electrons and two ions, but not all such quadruple detections result from DI due to false coincidences (e.g. two SI events). Additional filters are designed to check that the pair of ions add up to the parent ion and their momenta sum up to zero. Only about 0.2\% of the total coincidence events pass these filters. We plot the momentum correlations for both electrons and ions (CH$_2$Br$^+$ and I$^+$) collected in this fashion in Fig. \ref{fig:tpxdi}. We choose the laser polarization direction to be the momentum x-axis ($\mathrm{p_x}$). The bottom two ion-ion correlation plots show anti-correlations as expected from momentum conservation. It is interesting to note that the $\mathrm{p_x}$-$\mathrm{p_x}$ correlation has most yield for large momenta, while the $\mathrm{p_y}$-$\mathrm{p_y}$ correlation plot shows the largest yield around zero momentum. This implies that the ionization leading to these pair of fragment ions has the largest yield for molecules whose C-I axis parallel are aligned to the laser polarization.  The top two panels show the e-e correlations. We can see a diagonal line in both plots, which represent a positive correlation of the electrons' momenta - that is, both electrons have a tendency of being emitted in the same direction. This result is important because not only it shows the full quadruple coincidence VMI detection using the Tpx3Cam works as we expected, but also it provides evidence of electron correlation in molecular DI.

\section{Conclusions}
In conclusion, we have demonstrated the use of the new Tpx3Cam in conjunction with a velocity map imaging apparatus for coincidence detection of electrons and ions.  The high sensitivity and time resolution of Tpx3Cam allows for highly differential measurements with a single detector, and for switching between coincidence and non-coincidence modes easily and rapidly.  We anticipate that this configuration of VMI and Tpx3Cam will be useful for a wide range of experiments which produce electrons and ions whose momenta yield important information about molecular structure and dynamics.

\section{Acknowledgements}
We gratefully acknowledge support from the Department of Energy under award numbers DEFG02-08ER15983 and BNL LDRD grant 13-006 for the development of the apparatus. We gratefully acknowledge support from the National Science Foundation under award number 1205397 for carrying out the measurements with the apparatus.

\newpage
%

\end{document}